\documentclass[a4paper,11pt]{article}
\usepackage{jcappub} 
\usepackage{lineno}
\usepackage{amsmath,amssymb,amsfonts,mathtools,mathrsfs}
\usepackage{graphicx}
\usepackage{bm}
\usepackage{xcolor}
\usepackage{ulem,cancel,scalerel}
\usepackage{natbib}  
\usepackage[nameinlink,capitalise]{cleveref}
\usepackage{aas_macros}

\usepackage[scr]{rsfso}
\newcommand{\crefnp}[1]{{\crefformat{equation}{##2 Eq.\ ##1##3}\cref{#1}\crefformat{equation}{##2 Eq.\ (##1##3)}}}

\newcommand{\de}{\mathrm{d}}
\newcommand{\cH}{\mathcal{H}}
\newcommand{\ee}{\mathrm{e}}
\newcommand{\ii}{\mathrm{i}}

\newcommand{\para}{{\stretchrel*{\parallel}{\perp}}}

\newcommand{\xv}{\bm{x}}

\newcommand{\kv}{\bm{k}}
\newcommand{\nv}{\bm{n}}

\newcommand{\qv}{\bm{q}}

\newcommand{\Le}{{\mathcal L}}
\newcommand{\del}{\delta}
\newcommand{\X}{\rm{X}}
\newcommand{\Y}{\rm{Y}}

\let\Gamma\varGamma
\let\Delta\varDelta
\let\Theta\varTheta
\let\Lambda\varLambda
\let\Xi\varXi
\let\Pi\varPi
\let\Sigma\varSigma
\let\Upsilon\varUpsilon
\let\Phi\varPhi
\let\Psi\varPsi
\let\Omega\varOmega

\usepackage{wasysym}

\title{Complex yet Hermitian: Gaussian covariance of cross-correlation and multi-tracer power spectra}

\author[a,b,1]{Federico Montano,\note{Corresponding author.}}
\author[a,b,c,d]{Stefano Camera,}
\author[e,f,g]{Emiliano Sefusatti,}
\author[h,f,e,g]{Mohamed Yousry Elkhashab}

\affiliation[a]{Dipartimento di Fisica, Universit\`a degli Studi di Torino\\Via P.\ Giuria 1, 10125 Torino, Italy}
\affiliation[b]{INFN -- Istituto Nazionale di Fisica Nucleare, Sezione di Torino\\Via P.\ Giuria 1, 10125 Torino, Italy}
\affiliation[c]{INAF -- Istituto Nazionale di Astrofisica, Osservatorio Astrofisico di Torino\\Strada Osservatorio 20, 10025 Pino Torinese, Italy}
\affiliation[d]{Department of Physics \& Astronomy, University of Western Cape\\Cape Town 7535, South Africa}
\affiliation[e]{INAF -- Istituto Nazionale di Astrofisica, Osservatorio Astronomico di Trieste\\Via Tiepolo 11, 34131 Trieste, Italy}
\affiliation[f]{INFN -- Istituto Nazionale di Fisica Nucleare, Sezione di Trieste\\Via Valerio 2, 34127 Trieste, Italy}
\affiliation[g]{IFPU -- Institute for Fundamental Physics of the Universe\\Via Beirut 2, 34151 Trieste, Italy}
\affiliation[h]{Dipartimento di Fisica, Sezione di Astronomia, Universit\`a di Trieste\\Via Tiepolo 11, 34143 Trieste, Italy}

\emailAdd{federico.montano@unito.it}
\emailAdd{stefano.camera@unito.it}
\emailAdd{emiliano.sefusatti@inaf.it}
\emailAdd{mohamed.elkhashab@inaf.it}

\abstract{
Accurate modelling of the covariance of clustering observables is essential to fully exploit current and future survey data, which is expected to constrain large-scale clustering signals with unprecedented precision. 
Computational costs of simulation-based estimates motivate analytical approaches, especially in light of the growing interest towards multi-tracer analyses and parity-odd signatures in two-point statistics, which respectively mitigate cosmic variance and probe relativistic projection effects on cosmological scales.
In this work, we generalise previous theoretical results for the Gaussian covariance of multi-tracer power spectrum measurements, providing a general expression applicable to both real (even-parity) and complex (both even- and odd-parity) power spectra. 
We focus on a generic weighted estimator at first, and then showcase how our general formalism applies to Legendre power spectrum multipoles and two-dimensional power spectrum, recovering known limits in appropriate cases. We validate our predictions against Gaussian Monte Carlo simulations and investigate the structure of the covariance matrix, including the Hermitian properties of its imaginary part.
}

\begin{document}
\maketitle
\flushbottom

\section{Introduction}
\label{sec:intro}

Current and upcoming data require accurate methods to extract as much information as possible. Optical, infrared and radio surveys \cite{2025Euclid_OverviewA&A,2022_DESI_overview_AJ,2024_WST_whitepaper_arXiv,2022_MegaMapper_whitepaper_arXiv220904322S,2020_SKA_RedBook_PASA} are due to map the distribution of sources across unprecedented volumes, thus paving the way towards significant detections of large-scale physical phenomena, such as local primordial non-Gaussianity \cite{2004_Bartolo_PNG_PhR} and relativistic projection effects \cite{Yoo_2010,ChallinorLewis_2011,BonvinDurrer_2011}. 
Interest in such signals is responsible for ongoing efforts to properly account for physical and observational effects on the largest scales. Theoretical models have started including wide-angle effects \cite{2018_Castorina&White_MNRAS,Noorikuhani_2023}, together with novel signatures previously neglected. Amongst these is the anisotropic signal concealed in the imaginary part of the power spectrum, or, equivalently, in the odd multipoles of the two-point correlation function \cite{McDonald_2009,2020_Beutler&McDonald_JCAP,2021_Umeh_JCAP}. 

Multiple, upcoming surveys targeting the same area will allow to partially overcome the limits set by cosmic variance thanks to multi-tracers approaches \cite{2009_Seljak_PhRvL,2009_McDonald&Seljak_JCAP}.
It has been shown that joint analyses of summary statistics of different tracers of the (same, underlying) dark matter distribution are able to significantly tighten constraints on cosmological parameters, for each tracer can be interpreted as an independent realisation of the matter density field \cite{2013_Abramo&Leonard_MNRAS,2015_Fonseca_ApJ,2015_Alonso&Ferrera_PhRvD,borzyszkowski_liger_2017}. 
However, the size of multi-tracer data vectors is expected to be significant, therefore requiring large sets of mocks for robust numerical estimation of their covariance.

In this context, an accurate analytical modelling of the covariance matrix becomes particularly relevant. 
While avoiding the need for computationally expensive simulations, an analytical estimate is also unaffected by the noise characterising numerical ones \cite{2022_Percival2108.10402}. Furthermore, it can serve as a benchmark for de-noising techniques based on an eigenvalue decomposition \cite{2001_Eisenstein&Zaldarriaga,2005_Gaztanaga&Scoccimarro_MNRAS,farina2026denoisingclusteringcovariancematrices}, shrinkage \cite{Pope_Szapudi_08} or the simple fitting of a few parameters of a theoretical model to a poor numerical estimate  \cite{2016_OConnell_MNRAS,Pearson_2016,Friedrich_2017,2019_Hall_MNRAS,2021_Friedrich_MNRAS,2024_Euclid_FUmagalli}.
Building upon earlier works \cite{Feldman_1994, Scoccimarro_1999, Meiksin_1999, Hamilton_2006, 2016_Lacasa_JCAP,Grieb_2016, 2017_Howlett_MNRAS,2018_Lacasa_A&A,2020_Lacasa_A&A,2023_Bayer_PhRvD}, current analytical models for the covariance of two-point statistics in spectroscopic surveys now include non-Gaussian contributions as well as finite-volume effects \cite{Sugiyama_2020, Wadekar_2020, 2020_Philcox_MNRAS1904.11070,2020_Philcox_MNRAS1912.01010} and have been successfully adopted in the analysis of large datasets \cite{2020_Wadekar_PhRvD, 2025_Forero-Sanchez_JCAP}. 
The specific case of the covariance of cross-correlations was extensively explored, for the simple case of measurements in simulation boxes with periodic boundary conditions and real data points, in Ref.\ \cite{Smith_2009}. Subsequently, Refs.\ \cite{2018_Blake_MNRAS, Abramo_2022} showed how the cross-covariance could maximise the Fisher information, whereas Ref.\ \cite{Barreira_2020} analysed its use in optimising constraints on primordial non-Gaussianity \citep[see also][for an application to the \textit{Euclid} survey]{2024_Euclid_Multi_tracer_A&A}.

This paper stems from the twofold need to have covariance estimates which account for, on the one hand, possible odd-parity (hence imaginary) contributions to the galaxy power spectrum, and, on the other hand, multi-tracer analyses. We therefore focus on the Gaussian covariance of cross-correlation power spectrum estimators, and present general relations that apply both in the presence and absence of an imaginary component in the data vector, and consider an arbitrary number of tracers. 
Generality is ensured by writing the covariance for a class of estimators characterised by an unspecified weighting function of the angular dependence. Conversely, the desire to compare our results with other works and provide handy recipes for the most common estimators is then pursued by focusing on the Legendre multipoles of the power spectrum and the full two-dimensional (2D)---namely, binned according to the wavenumber and the cosine of the wavevector w.r.t. the line-of-sight LOS---power spectrum. 
Moreover, the robustness of our results is checked by comparing them against covariance estimates from Gaussian realisations of the density field.
Following Ref.\ \cite{Grieb_2016}, our key assumptions are Gaussianity and a trivial survey geometry. The former ensures that the power spectrum provides a complete description of the statistics of the matter density field. The latter---a uniform, cubic volume with periodic boundary conditions---makes Fourier space algebra tractable and implies that neither window convolution nor beyond plane-parallel contributions are considered.

We structure this paper as follows. In \cref{sec:conventions}, we establish our notation and Fourier convention; in \cref{sec:estimators}, we first define a general, cross-power spectrum estimator and then recast it in the form of Legendre multipoles and full-2D power spectrum, our case studies. \Cref{sec:covariance} conveys the key results of our work, presenting the general Gaussian covariance matrix associated with multi-tracer measurements of the power spectrum. Then, in \cref{sec:validation}, we validate the accuracy of our calculations for both the power spectrum multipoles and the 2D power spectrum against simple Gaussian realisations. Conclusions are drawn in \cref{sec:conclusion}, while further details about the derivations of the covariance are left in \cref{app:derivation_covPell,app:cov_imaginary_part}.

\section{Definitions}
\label{sec:conventions}

\subsection{Fourier conventions}
We adopt the following Fourier-transform convention:
\begin{equation}
f(\kv) = \int_V \de^3x\, \ee^{-\ii\kv\cdot\xv}\,f(\xv)\,,
\end{equation}
with inverse
\begin{equation}
f(\xv) = \frac{1}{V}\sum_{\kv} \ee^{\ii\kv\cdot\xv}\,f(\kv)\,,
\end{equation}
written here for a generic function $f$ whose argument distinguishes between Fourier transform and anti-transform, and with $V=L^3$ representing a cubic volume with periodic boundary conditions. 
We therefore explicitly assume a finite volume that leads to discrete wavevectors $\kv = (2\,\pi/L)\,\nv$, with $\nv\in\mathbb{Z}^3$, and define the fundamental frequency as $k_{\rm f}=2\,\pi/L$. This sets the largest physical wavelength supported by the box and thereby determines the spacing between neighbouring Fourier modes in the discretised volume.
In our convention, the Dirac-delta function in configuration space and the Kronecker-delta symbol in Fourier space can be respectively written as
\begin{equation}
\delta^{\rm D}(\xv) = \frac{1}{V}\sum_{\kv} \ee^{\ii\kv\cdot\xv}
\qquad {\rm and} \qquad
\delta^{\rm K}_{\kv} = \frac{1}{V}\int \de^3x\, \ee^{-\ii\kv\cdot\xv}\,,
\end{equation}
where $\delta^{\rm K}_{\kv} \coloneqq\delta^{\rm K}_{\kv,\,0}=1$ for $\kv=0$ and vanishes otherwise. For real fields, the Fourier transforms satisfy the reality condition, $f^\ast(\kv)=f(-\kv)$.

\subsection{Power spectrum}

Given the density contrasts  $\delta_X$ and $\delta_Y$, of two tracers of the large-scale structure (LSS), $X$ and $Y$, their two-point correlation function in Fourier space, i.e.\ the power spectrum \(P_{XY}(\kv)\), is implicitly defined via
\begin{equation}
\label{eq:ps}
\langle \delta_X(\kv)\,\delta_Y(\kv')\rangle \eqqcolon (2\,\pi)^3\,\frac{\delta^{\rm K}_{\kv,-\kv'}}{k_{\rm f}^3}\,P_{XY}(\kv)=V\,\del^K_{\kv,-\kv'}\,P_{XY}(\kv)\,,
\end{equation}
where $\langle\,\dots\rangle$ denotes ensemble average.
\Cref{eq:ps} reduces to the well-known auto-spectrum expression if $X=Y$, and refers to a cross-spectrum for \(X\ne Y\).
In our convention, the infinite-volume limit, $V\rightarrow \infty$, can be recovered by replacing $\delta^{\rm K}_{\kv}/k_{\rm f}^3\rightarrow \delta^{\rm D}(\kv)$, such that
\begin{equation}
\label{eq:ps_continuous}
\langle \delta_X(\kv)\,\delta_Y(\kv')\rangle \equiv (2\,\pi)^3\,\delta^{\rm D}(\kv+\kv')\,P_{XY}(\kv)\,.
\end{equation}

We assume redshift-space observables in the plane-parallel approximation, so that $P_{XY}$ is anisotropic and depends on the vector $\kv$. 
It may be worth noticing that, in general, $P_{XY}(\kv)\neq P_{YX}(\kv)$, since peculiar velocities can induce odd-parity, imaginary contributions in cross-power spectra; rather, $P_{XY}(\kv)= P_{YX}(-\kv)$ holds as a result of the reality condition on $\delta_X(\kv)$.
Statistical symmetries always imply invariance under rotations about the LOS, so that the dependence reduces to two variables.
A convenient parametrisation is obtained by decomposing $\kv$ into components perpendicular and parallel to the LOS, $\kv = (\kv_\perp, k_\para)$, yielding the so-called cylindrical power spectrum, $P_{XY}(k_\perp, k_\para)$. Equivalently, one can write $P_{XY}(\kv) = P_{XY}(k,\mu)$, where $\mu = k_\para/k$.
It is often useful to expand the angular dependence in Legendre polynomials, defining the multipoles as
\begin{equation} \label{eq:multipole_exp}
P_{XY}(k,\mu) = \sum_{\ell=0}^\infty P_{XY,\ell}(k)\,\Le_\ell(\mu)\,.
\end{equation}

To conclude this section, we point out that none of the key calculations of this paper will strictly require the density contrast fields ($\delta_X,\, \delta_Y,\ldots$) to come from a galaxy catalogue. We, in fact, shall refer to galaxy clustering observations when considering sampling noise---assumed Poissonian---and validating our predictions; however, this has to be intended as nothing but a convenient choice to assess the correctness of our derivations. Power spectra can refer to any kind of tracer, provided one knows how to model them and their associated noise.

\section{Estimators}
\label{sec:estimators}

We consider the following estimator for the cross-correlation power spectrum
\begin{align}
\label{eq:estP}
\hat{P}_{XY,\,\alpha}(k)  & \coloneqq   \frac{k_{\rm f}^3}{N_k}\sum_{\qv \in k} \,w_\alpha(\qv)\,\delta_X(\qv)\,\delta_Y(-\qv)\,\\
& \stackrel{V\rightarrow \infty}{=} \frac{k_{\rm f}^3}{V_k}\int_{k}\de^3q \,w_\alpha(\qv)\,\delta_X(\qv)\,\delta_Y(-\qv)\,,
\label{eq:estP_cont}
\end{align}
where the sum runs over all the $N_k$ modes, $\{\qv\}$, in a spherical shell of radius $k$ and width $\Delta k$, so that
\begin{equation}
\label{eq:Nk}
N_k  \coloneqq \sum_{\qv \in k}= \frac{V_k}{k_f^3} \simeq  \frac1{k_f^3} \int_{k-\Delta k/2}^{k+\Delta k/2} \de \, q\,q^2 \, \de\Omega_{\qv}
 =  4\pi \, \frac{k^2\, \Delta k}{k_f^3} + {\mathcal O}(\Delta k^3)\,.
\end{equation}
The function $w_\alpha$ is introduced to write a compact expression for a generic estimator. In fact, the spherical averaged power spectrum, i.e.\ the monopole, will then correspond to $w_{\alpha}(\qv)\coloneqq1$. More generally, for
\begin{equation}
w_{\ell}(\qv)\coloneqq(2\,\ell+1)\, \Le_{\ell}(\mu_{\qv})\,,
\end{equation}
with $\mu_{\qv}$ the cosine of the angle between the wavevector and the LOS, one obtains the usual multipoles estimator:
\begin{equation}
\label{eq:estPell}
    \hat{P}_{\X\Y,\ell}(k)  \coloneqq  (2\,\ell+1)\,\frac{k_f^3}{N_k}\sum_{\qv \in k} \, \Le_\ell({\mu_{\qv}})\delta_{\X}(\qv)\,\delta_{\Y}(-\qv)\,.
\end{equation}
On the other hand, assuming
\begin{equation}
w_{\mu}(\qv)\coloneqq
\frac{2}{\Delta\mu}\,[\Theta(\mu_{\qv} -\mu+\Delta\mu/2)\,\Theta(\mu+\Delta\mu/2-\mu_{\qv})]\,,
\end{equation}
with $\Theta$ the Heaviside step function, the basic estimator for the 2D power spectrum is defined in terms of a bin for $\mu_{\qv}$, centred on $\mu$ and of size $\Delta \mu$,
\begin{align}
    \hat{P}_{XY}(k,\mu) & \coloneqq  
      \frac{k_{\rm f}^3}{N_{k}}\sum_{\qv \in k}  \frac{2}{\Delta\mu}\,\left[\Theta\left(\mu_{\qv} -\mu+\Delta\mu/2\right)\,\Theta\left(\mu+\Delta\mu/2-\mu_{\qv}\right)\right] \, \delta_X(\qv)\,\delta_Y(-\qv) \nonumber \\ & = 
        \frac{k_{\rm f}^3}{N_k}\,\frac{2}{\Delta\mu}\,\sum_{q=k-\Delta k/2}^{k+\Delta k/2}\sum_{\mu_{\qv}=\mu-\Delta\mu/2}^{\mu+\Delta\mu/2} \delta_X(\qv)\,\delta_Y(-\qv)\nonumber \\
        &\equiv\frac{k_{\rm f}^3}{N_{k,\mu}}\,\sum_{\qv \in k,\mu} \delta_X(\qv)\,\delta_Y(-\qv)\,,\label{eq:estPkmu}
\end{align}
where $N_{k,\mu}=N_k \, \frac{\Delta\mu}{2}$, and the last sum is restricted to $k-\Delta k/2\le q < k+\Delta k/2$ and $\mu-\Delta\mu/2\le \mu_{\qv} < \mu+\Delta\mu/2$.

We notice that, in presence of odd-parity terms, $\hat{P}_{XY}(k,\mu)$ receives different contributions from the modes $\qv$ and $-\qv$. This motivates a binning scheme allowing for positive and negative values of $\mu$, as opposed to the usual case of clustering wedges\footnote{Clustering wedges are in fact defined as bins in the modulus of the cosine of the angle the wavevector forms with the LOS.} (for comparison, see \cref{tab:estimators_w}).
\begin{table}
    \centering
    \begin{tabular}{ll}
        Estimator & $w_\alpha(\qv)$ \\ \hline
        Legendre Multipoles & $(2\,\ell+1)\, \Le_{\ell}(\mu_{\qv})$ \\
        Full-2D power spectrum & $2\,[\Theta(\mu_{\qv} -\mu+\Delta\mu/2)\,\Theta(\mu+\Delta\mu/2-\mu_{\qv})]/\Delta\mu$ \\
        Clustering wedges & $[\Theta(|\mu_{\qv}| -\mu+\Delta\mu/2)\,\Theta(\mu+\Delta\mu/2-|\mu_{\qv}|)]/\Delta\mu$ \\
    \end{tabular}
    \caption{Definitions of $w_\alpha(\qv)$ for the main estimators that can be written in the form of \cref{eq:estP}.}
    \label{tab:estimators_w}
\end{table}

\section{Covariance}
\label{sec:covariance}

\subsection{General expression}
The general expression for the power spectrum estimator in \cref{eq:estP}, encompassing both redshift-space multipoles as well as wedges, allows to compute the cross-covariance between generic cross-correlations in a general form. This is defined, for four distinct tracers, as
\begin{equation}\label{eq:cov_def}
    \mathsf{C}[\hat{P}_{XY,\,\alpha}(k),\,\hat{P}_{ZW,\,\beta}(k')  ] \coloneqq
    \left\langle\delta\hat{P}_{XY,\,\alpha}(k)\,\delta\hat{P}_{ZW,\,\beta}^\ast(k')\right\rangle\,,
\end{equation}
where \(\delta\hat{P}_{XY,\,\alpha}\coloneqq\hat{P}_{XY,\,\alpha}-\langle\hat{P}_{XY,\,\alpha}\rangle\).
The complex conjugate on the second term ensures the covariance matrix to be Hermitian and the corresponding variance positive definite.

The definition (\crefnp{eq:cov_def}) expands into
\begin{align}
    \mathsf{C}[\hat{P}_{XY,\,\alpha}(k),\,\hat{P}_{ZW,\,\beta}(k')  ] = &
    \frac{k_{\rm f}^6}{N_k\, N_{k'}}\sum_{\qv \in k} \sum_{\qv' \in k'} \,w_\alpha(\qv)\,w_\beta(\qv')\,
    \langle\delta_X(\qv)\,\delta_Y(-\qv)\delta_Z^\ast(\qv')\,\delta_W^\ast(-\qv')\rangle \nonumber \\ &
    - \langle\hat{P}_{XY,\,\alpha}(k)\rangle\,\langle\hat{P}^\ast_{ZW,\,\beta}(k')\rangle\,, \label{eq:cov_der1}
\end{align}
where $\langle\ldots\rangle$ is the four-point correlation function in Fourier space, which we evaluate under two relevant assumptions: Gaussianity and independent tracers \citep[see][for comparison]{Smith_2009}.
The latter means that, given for instance $X$ and $Y$, they can either be the same tracer ($X=Y$) or two non-overlapping populations that track the same underlying matter field ($X \neq Y$). The former allows us to use Wick's theorem and neglect the connected part,
\begin{align}
    \langle\delta_X(\qv)\,\delta_Y(-\qv)\delta_Z^\ast(\qv')\,\delta_W^\ast(-\qv')\rangle = & \,
    \langle\delta_X(\qv)\,\delta_Y(-\qv)\rangle\,\langle\delta_Z^\ast(\qv')\,\delta_W^\ast(-\qv')\rangle  \nonumber \\ &\,  + 
    \langle\delta_X(\qv)\,\delta_Z^\ast(\qv')\rangle\,\langle\delta_Y(-\qv)\,\delta_W^\ast(-\qv')\rangle 
    \nonumber \\ &\,  +
    \langle\delta_X(\qv)\,\delta_W^\ast(-\qv')\rangle\,\langle\delta_Y(-\qv)\,\delta_Z^\ast(\qv')\rangle\,.
\end{align}
The contribution of the first term above (first line) to \cref{eq:cov_der1} gives $\langle\hat{P}_{XY,\,\alpha}(k)\rangle\langle\hat{P}^\ast_{ZW,\,\beta}(k')\rangle$, hence cancels out with that in the second line.
Further, applying \cref{eq:ps} we find
\begin{align}
    \mathsf{C}[\hat{P}_{XY,\,\alpha}(k),\,\hat{P}_{ZW,\,\beta}(k')  ] = &
    \frac{1}{N_k\, N_{k'}}\sum_{\qv \in k} \sum_{\qv' \in k'} \,w_\alpha(\qv)\,w_\beta(\qv') 
    \nonumber \\ & \times 
    \left[\delta^{\rm K}_{\qv,\qv'}\,P_{XZ}(\qv)\,\delta^{\rm K}_{\qv,\qv'}\,P_{YW}(-\qv)
    + \delta^{\rm K}_{\qv,-\qv'}\,P_{XW}(\qv)\,\delta^{\rm K}_{\qv,-\qv'}\,P_{YZ}(-\qv)\right]
    \nonumber \\ = &
    \frac{\delta^{\rm K}_{k,k'}}{N_k^2}\sum_{\qv \in k} \,w_\alpha(\qv) 
    \nonumber \\ & \times 
    \left[w_\beta(\qv)\, P_{XZ}(\qv)\,P_{YW}(-\qv)
    + w_\beta(-\qv)\, P_{XW}(\qv)\,P_{YZ}(-\qv)\right]\,,
\end{align}
and, using the reality condition on the power spectra,\footnote{The reality condition applies not only to the density contrast $\del_X(\kv)$, but also on the galaxy power spectrum, as this is the Fourier counterpart of the two-point correlation function, a real quantity itself.}
\begin{multline} \label{eq:covPs_general}
    \mathsf{C}[\hat{P}_{XY,\,\alpha}(k),\,\hat{P}_{ZW,\,\beta}(k')  ] = \\
    \frac{\delta^{\rm K}_{k,k'}}{N_k^2}\sum_{\qv \in k} \,w_\alpha(\qv) 
    \,\left[w_\beta(\qv)\, P_{XZ}(\qv)\,P_{WY}(\qv)
    + w_\beta(-\qv)\, P_{XW}(\qv)\,P_{ZY}(\qv)\right]\,.
\end{multline}

\Cref{eq:covPs_general} is the main general result of this paper. It explicitly reflects the two relevant couplings between the Fourier modes: the variance contribution, i.e.\ $\qv=\qv'$, and $\qv=-\qv'$, due to the reality condition. 
The covariance matrix is Hermitian both w.r.t. the wavevector and the tracers, since the power spectrum obeys $P_{XY}^\ast(\qv)=P_{XY}(-\qv)=P_{YX}(\qv)$.

\subsubsection{Shot noise}
In galaxy surveys, the observed density contrast field includes not only the underlying cosmological signal but also a stochastic noise contribution due to the discrete sampling of galaxies. For a given sample $X$, the observed Fourier-space density contrast can be written as
\begin{equation}\label{eq:observed_filed_w_noise}
\delta_X^{\rm obs}(\bm k) = \delta_X(\bm k) + \epsilon_X(\bm k) \,,
\end{equation}
where $\delta_X(\bm k)$ is the true cosmological signal and $\epsilon_X(\bm k)$ is a noise term. Assuming Poisson sampling, the noise satisfies
\begin{equation}
\langle \epsilon_X(\kv) \epsilon_Y(\kv') \rangle =(2\,\pi)^3 \,\frac{\delta^{\rm K}_{\kv,-\kv'}}{k_{\rm f}^3}  \frac{\delta^{\rm K}_{X,Y}}{\bar n_X} , \qquad
\langle \delta_X(\kv) \epsilon_Y(\kv') \rangle = 0 \,.
\end{equation}
As a consequence, we can account for shot noise in the covariance of galaxy clustering measurements via the substitution 
\begin{equation} \label{eq:noise}
P_{XY}(\bm k) \longrightarrow P_{XY}(\bm k) + \frac{\delta^{\rm K}_{X,Y}}{\bar n_X}
\end{equation}
in \cref{eq:covPs_general}, so that shot noise contributes a scale-independent term only to the auto-power spectrum.

\subsubsection{Auto- and cross-correlation cases}
If $X=Y=Z=W$, \cref{eq:covPs_general} reduces to
\begin{equation} \label{eq:Cov_modebymode_auto}
    \mathsf{C}[\hat{P}_{XX,\,\alpha}(k),\,\hat{P}_{XX,\,\beta}(k')  ] = 
    \frac{\delta^{\rm K}_{k,k'}}{N_k^2}\sum_{\qv \in k} \,w_\alpha(\qv) \, P_{XX}^2(\qv)
    \,\left[w_\beta(\qv)+ w_\beta(-\qv)\right]\,,
\end{equation}
that is the well-known expression of the auto-power spectrum covariance, rephrased in terms of a general estimator in the form of \cref{eq:estP}.
This corresponds to the Gaussian limit of, for instance, Eq.\ (5) in Ref.\ \cite{Meiksin_1999} and Eq.\ (11) in Ref.\ \cite{2024_Euclid_FUmagalli}.

On the other hand, if $X=Z$ and $Y=W$, \cref{eq:covPs_general} becomes
\begin{equation}\label{eq:Cov_gen_cross}
    \mathsf{C}[\hat{P}_{XY,\,\alpha}(k),\,\hat{P}_{XY,\,\beta}(k')  ] = 
    \frac{\delta^{\rm K}_{k,k'}}{N_k^2}\sum_{\qv \in k} \,w_\alpha(\qv) 
    \,\left[w_\beta(\qv)\, P_{XX}(\qv)\,P_{YY}(\qv)
    + w_\beta(-\qv)\, P_{XY}^2(\qv)\right]\,.
\end{equation} 
This is the variance of the cross-power spectrum $P_{XY}$, explicitly Hermitian under $X\longleftrightarrow Y$.
In the real-space case, \cref{eq:Cov_gen_cross} corresponds to the Gaussian limit of Eq.\ (48) in Ref.\ \cite{Smith_2009}.

\subsection{Power spectrum multipoles}
We obtain the case of cross-covariance of redshift-space multipoles of cross-power spectra (already defined in \cref{eq:estPell}) by substituting $w_{\ell}(\qv)\coloneqq(2\,\ell+1)\, \Le_{\ell}(\mu_{\qv})$ into our general expression.
Denoting the multipoles of interest by $\ell$ and $\ell'$, \cref{eq:covPs_general} yields
\begin{align}
    \mathsf{C}[\hat{P}_{XY,\,\ell}(k),\,\hat{P}_{ZW,\,\ell'}(k')  ] &= 
    \frac{(2\,\ell+1)\,(2\,\ell'+1)\,\delta^{\rm K}_{k,k'}}{N_k^2} \nonumber \\&
    \times\sum_{\qv \in k} \,\Le_\ell(\mu_{\qv}) \, \Le_{\ell'}(\mu_{\qv})
    \,\left[P_{XZ}(\qv)\,P_{WY}(\qv)
    + (-1)^{\ell'}\, P_{XW}(\qv)\,P_{ZY}(\qv)\right]\,, 
    \label{eeq:cov_Pell_start}
\end{align}
where the parity of Legendre polynomials, $\Le_\ell(\mu)=(-1)^\ell \Le_\ell(-\mu)$, has been exploited. 

To understand how the different multipoles contribute to the covariance, we use \cref{eq:multipole_exp} and expand the power spectra on the right-hand side in multipoles too, then study the properties of the integral (in the continuum limit) over four Legendre polynomials we will be left with, as in Refs.\ \cite{Beutler_2020,Grieb_2016}. 
For the sake of readability, we report the final result below and leave the detailed derivation in Appendix \ref{app:derivation_covPell}.

We find the covariance:
\begin{align}
    \mathrm{Cov}({\hat P_{XY,\ell}}({k}),&\, {\hat P_{ZW,\ell'}}({ k'})) = \frac{2\,\pi\,\delta^{\rm K}_{k,k'}}{N^2_k \, k_{\rm f}^3} \,  
    (2\,\ell+1)\,(2\,\ell'+1)
    \nonumber\\
    &\quad\times\sum _{L_1,L_2,L_3} \, (2\,L_3+1) \,
    \begin{pmatrix} 
        \ell & \ell' & L_3 \\ 
        0 & 0 & 0 
    \end{pmatrix}^2
    \begin{pmatrix} 
        L_1 & L_2 & L_3 \\ 
        0 & 0 & 0 
    \end{pmatrix}^2
    \nonumber\\
    &\qquad\times
    \int_{k-\Delta k/2}^{k+\Delta k/2}
    \de q\, q^2\,\Big\{P_{XZ,L_1}(q) \, P_{WY,L_2}(q) + (-1)^{\ell'}\, P_{XW,L_1}(q)\, P_{ZY,L_2}(q) \Big\} \,,\label{eq:covariance_Pl_full_multi}
\end{align}
where round brackets denote Wigner-3j symbols, capitalised multipole indexes are summed over, and we have translated to the continuum limit, according to \cref{eq:estP_cont}, in order to take advantage of the exact angle integrations. For galaxy clustering observations, as usual, Poissonian shot noise only affects monopole contributions.
\Cref{eq:covariance_Pl_full_multi} is the second main result of our paper and shows that the Gaussian cross-covariance matrix mixes different multipole and tracer contributions. 

Key selection rules are provided by
\begin{equation*}
    \begin{pmatrix} \ell_1 & \ell_2 & \ell_3 \\ 0 & 0 & 0  \end{pmatrix}\neq 0 \quad \textrm{ only if } \quad 
    (\ell_1+\ell_2+\ell_3)/2\in \mathbb{N} \; \textrm{ and } \; |\ell_1-\ell_2|\le\ell_3\le\ell_1+\ell_2\,,
\end{equation*}
and have relevant consequences.
First, the variance ($\ell=\ell'$) or the covariance of same parity multipoles ($(\ell+\ell')/2\in \mathbb{N}$) always force $(L_1+L_2)/2\in \mathbb{N}$---that is, individually, every term of the sum does not mix odd and even multipoles. An interesting implication of this is that the shot noise is expected to enter the variance of all the multipoles.
Secondly, the covariance of different parity multipoles ($\ell$ even and $\ell'$ odd, or vice versa) only gets contributions from $(L_1+L_2+1)/2\in \mathbb{N}$, i.e.\ it forces $L_1$ and $L_2$ to have opposite parity.
Thirdly, let us remind that odd(even) multipoles are only imaginary(real) by construction and notice that the covariance matrix in \cref{eq:covariance_Pl_full_multi} is Hermitian under $\ell\longleftrightarrow\ell'$ as it is ought to be.

\subsubsection{Notable cases}
In the case of a spherical-averaged power spectrum, which corresponds to the monopole of Legendre polynomials (as well as the single \(\mu\)-bin case, \(\Delta\mu=2\), of the 2D power spectrum), \cref{eeq:cov_Pell_start,eq:covariance_Pl_full_multi} reduce to
\begin{equation}
    \mathrm{Cov}[{\hat P_{XY,0}}({k}), {\hat P_{ZW,0}}({ k'})] =
    \frac{2\,\pi\,\delta^{\rm K}_{k,k'}}{N^2_k \, k_{\rm f}^3}  
    \int_{k-\Delta k/2}^{k+\Delta k/2}
    \de q\, q^2\,\Big\{P_{XZ,0}(q) \, P_{WY,0}(q) + P_{XW,0}(q)\, P_{ZY,0}(q) \Big\} \,,\label{eq:covariance_monopole}
\end{equation}
which, in the limit of auto-correlation ($X=Y=Z=W$), is nothing but the so-called variance of the FKP estimator \cite{Feldman_1994} \citep[see also][]{Sefusatti_2006, DODELSON_SCHMIDT_cap14_2025}.
As a consequence of Wigner-3j symbol properties, only the monopole contributes to the monopole covariance. Also, when $X=Z$ and $Y=W$, \cref{eq:covariance_monopole} is consistent with Refs.\ \cite{Abramo_2022,Barreira_2020}.

Moreover, for cross-power spectra \cref{eeq:cov_Pell_start} becomes
\begin{align}
    \mathsf{C}[\hat{P}_{XY,\,\ell}(k),\,\hat{P}_{XY,\,\ell'}(k')  ] &= 
    \frac{(2\,\ell+1)\,(2\,\ell'+1)\,\delta^{\rm K}_{k,k'}}{N_k^2} \nonumber \\&
    \times\sum_{\qv \in k} \,\Le_\ell(\mu_{\qv}) \, \Le_{\ell'}(\mu_{\qv})
    \,\left[P_{XX}(\qv)\,P_{YY}(\qv)
    + (-1)^{\ell'}\, P_{XY}^2(\qv)\right]\,,
\end{align}
in line with Ref.\ \cite{Addis_2025}. Such a relation explicitly shows that, when $\ell$ and $\ell'$ are both odd, we essentially extract the imaginary part of the power spectrum (plus some noise terms; we refer the reader to \cref{app:cov_imaginary_part} for further details), and are not left with a zero-variance term (because in general $P_{XY}^2(\qv)\neq|P_{XY}(\qv)|^2$).

Another consequence of our multi-tracer result is its auto-correlation limit,
\begin{multline}
    \mathsf{C}[{\hat P_{XX,\ell}}({k}), {\hat P_{XX,\ell'}}({ k'})] = \frac{2\,\pi\,\delta^{\rm K}_{k,k'}}{N^2_k \, k_{\rm f}^3} \, 
     (2\,\ell+1)\,(2\,\ell'+1)\\
    \times\sum _{L_1,L_2,L_3} \, (2\,L_3+1)
    \begin{pmatrix} 
        \ell & \ell' & L_3 \\ 
        0 & 0 & 0 
    \end{pmatrix}^2
    \begin{pmatrix} 
        L_1 & L_2 & L_3 \\ 
        0 & 0 & 0 
    \end{pmatrix}^2\\
    \times\int_{k-\Delta k/2}^{k+\Delta k/2}
    \de q\, q^2\,\left[1 + (-1)^{\ell'}\right] P_{XX,L_1}(q)\, P_{XX,L_2}(q) \,,\label{eq:covariance_Pl_full_auto}
\end{multline}
whose prediction for any terms involving odd multipoles is identically zero, thus reflecting the auto-power spectrum being always a real quantity.

\subsection{Full-2D power spectrum}
Moving to the second selected estimator, the full two-dimensional power spectrum, we apply \cref{eq:estPkmu} to \cref{eq:covPs_general} and write:
\begin{equation} \label{eq:cov_Pkmu_multi}
    \mathsf{C}[\hat{P}_{XY}(k,\mu),\,\hat{P}_{ZW}(k',\mu')  ] = 
    \frac{\delta^{\rm K}_{k,k'}}{N_{k,\mu}^2}\sum_{\qv \in k,\mu}
    \,\left[\delta^{\rm K}_{\mu,\mu'} \, P_{XZ}(\qv)\,P_{WY}(\qv)
    + \delta^{\rm K}_{\mu,-\mu'} \, P_{XW}(\qv)\,P_{ZY}(\qv)\right]\,,
\end{equation}
where we have used a symmetric binning in $\mu_{\qv}$, with fixed $\Delta\mu$.

It is worth noticing some consequences of \cref{eq:cov_Pkmu_multi}. 
The covariance matrix is diagonal in $k$, but it is only block-diagonal in $\mu$, due to the coupling of the ${\qv}=(q,\mu_{\qv})$ mode to the ${-\qv}=(q,-\mu_{\qv})$ one, given by the reality condition; this results in imaginary contributions affecting the off-diagonal part.
Moreover, cross-correlations play a crucial role when this estimator is adopted. Indeed, only when the off-diagonal contributions are different from the diagonals' is \cref{eq:cov_Pkmu_multi} not a singular matrix. This means that the auto-power spectrum covariance matrix $\mathsf{C}[\hat{P}_{XX}(k,\mu),\,\hat{P}_{XX}(k',\mu')]$ happens to be singular, therefore this estimator is not usable in the case of auto-correlations and the natural solution is to restore to the well-known definition of the clustering wedges (in which case \cref{eq:cov_Pkmu_multi} would become diagonal instead).
If $X=Z$ and $Y=W$, we have
\begin{equation}
\mathsf{C}[\hat{P}_{XY}(k,\mu),\,\hat{P}_{XY}(k',\mu')]
=  
\frac{\delta^{\rm K}_{k,k'}}{N_{k,\mu}^2}
\sum_{\qv \in k,\mu}
\,\left[\delta^{\rm K}_{\mu,\mu'} \, P_{XX}(\qv)\,P_{YY}(\qv)
+ \delta^{\rm K}_{\mu,-\mu'} \, P_{XY}^2(\qv)\right]\,,
\end{equation}
which can be arranged in a block-diagonal matrix, with each---Hermitian as it must be---block given by
\begin{multline}
    \begin{pmatrix}
        \mathsf{C}[\hat{P}_{XY}(k,\mu),\,\hat{P}_{XY}(k',\mu')] &
        \mathsf{C}[\hat{P}_{XY}(k,\mu),\,\hat{P}_{XY}(k',-\mu')] \\
        \mathsf{C}[\hat{P}_{XY}(k,-\mu),\,\hat{P}_{XY}(k',\mu')] &
        \mathsf{C}[\hat{P}_{XY}(k,-\mu),\,\hat{P}_{XY}(k',-\mu')]
    \end{pmatrix} 
    = \\ 
    \frac{\delta^{\rm K}_{k,k'}}{N_{k,\mu}^2}\,
    \begin{pmatrix}
        \sum_{\qv \in k,\mu}  P_{XX}(q,\mu_{\qv})\,P_{YY}(q,\mu_{\qv}) &
        \sum_{\qv \in k,\mu}  P_{XY}^2(q,\mu_{\qv})\\
        \sum_{\qv \in k,\mu}  P_{XY}^2(q,-\mu_{\qv})&
        \sum_{\qv \in k,\mu}  P_{XX}(q,-\mu_{\qv})\,P_{YY}(q,-\mu_{\qv})
    \end{pmatrix}\,. \label{eq:cov_Pkmu_block_cross}
\end{multline}

We point out that \cref{eq:cov_Pkmu_multi} is not in complete agreement with the usual formulation of the multi-tracer covariance matrix present in the literature. For example, eqs.\ (18-23) in Ref.\ \cite{White_2009}  (almost implicitly), eq.\ (3) in Ref.\ \cite{Zhao_2021}, eq.\ (3.3) in Ref.\ \cite{Dlamini_2024}, eq.\ (2.5) in Ref.\ \cite{Barberi_Squarotti_2024}, eq.\ (28) in Ref.\ \cite{Jolicoeur_2023}, and eq.\ (4.3) in Ref.\ \cite{Kopana_2024} all point to a diagonal covariance matrix in $\mu$, that resembles both contributions in \cref{eq:cov_Pkmu_multi} into the term proportional to $\delta^{\rm K}_{\mu,\mu'}$---that is, misinterpreting the $(\mu, -\mu)$ contribution. Along the same line are Refs.\ \cite{Montano_2023} and \cite{Montano_2024}, which extend the formalism of such a diagonal covariance matrix to complex power spectra. A fairly similar case is that of Ref.\ \cite{Karagiannis_2024}, where the authors 
look at the monopole---in consistency with Ref.\ \cite{Barreira_2020}---though include RSDs, not making explicit the integration over the Fourier bins. 

\section{Validation}
\label{sec:validation}
We now move to the numerical validation of our analytical prediction, filling this Section with further information about the form and accuracy of the covariance matrices we have derived in \cref{sec:covariance}.
The correctness of our estimates is tested against Monte Carlo mock measurements of the power spectra: we generate $5\,000$ Gaussian realisations in Fourier space of the matter density field and look at the covariance matrices associated with estimates of $\hat{P}_{XY,\ell}(k)$ and $\hat{P}_{XY}(k,\mu)$.
Relying on simulations that match our assumptions, we offer insights about the robustness of our theoretical predictions and their implications when it comes to multi-tracer analyses of (possibly complex) power spectra.

We adopt a linear model for the density contrast of the tracers, so that $\delta_X(\kv)=T_X(\kv)\,\delta_{\rm m}(\kv)$ and $\langle\delta_{\rm m}(\kv)\,\delta_{\rm m}(\kv')\rangle \coloneqq (2\,\pi)^3\,k_{\rm f}^{-3}\,\delta^{\rm K}_{\kv,-\kv'}\,P_{\rm m}(\kv)$, with $\delta_{\rm m}$ and $P_{\rm m}$, respectively, the matter density contrast and (linear) power spectrum, evaluated at the considered redshift. 
As customary in the case of galaxy clustering observations, we include shot noise in our simulations, following \cref{eq:observed_filed_w_noise}, and subtract it in the mock measurements.
Given our interest in treating odd-parity contributions, we adopt, for both simulations and evaluation of the theoretical covariance in \cref{eq:covPs_general}, a simple model for the transfer functions that includes the relativistic Doppler term, on top of the standard (linear bias plus redshift-space distortions) kernel \cite{Yoo_2010,ChallinorLewis_2011,BonvinDurrer_2011,2017AbramoBertacca1706.01834}. We thus use
\begin{equation} \label{eq:T_delta_g_Fourier_space}
    T_X(k,\mu)=b_X\,+f\,\mu^2+\ii\,\frac{\cH}{k}\,\alpha_X\,f\,\mu\,,
\end{equation}
with $b$ the galaxy bias, $f$ the growth factor, $\cH$ the conformal Hubble factor, and $\alpha$ the amplitude of the Doppler corrections, as in Ref.\ \cite{Montano_2023}.
To ensure reproducibility of our numerical validation, we leave in \cref{app:validation_details} information about redshift and specific values of parameters we have used. 

We look at the covariance associated with a measurement of a cross-correlation spectrum (i.e.\ $X=Z$ and $Y=W$) since it is the simplest, non-trivial observable which might include imaginary contributions. The auto-correlation is hence realised in the, always real-only, $X=Y$ limit.  Also, the properties of a full multi-tracer covariance matrix, where a larger data vector containing multiple spectra is considered, can be inferred on a block-by-block basis thanks to the generality of our four-tracer formulation in \cref{sec:covariance}, as, for instance, $\hat{\bm P} = (\hat{P}_{XY},\hat{P}_{ZW},\ldots)$ implies
\begin{equation} 
     \mathsf{C}[\hat{\bm P},\,\hat{\bm P} ] = 
    \left(
    \begin{array}{ccc}
    \mathsf{C}[\hat{P}_{XY}\,\hat{P}_{XY}]    & \mathsf{C}[\hat{P}_{XY}\,\hat{P}_{ZW}]   & \ldots   \\
    \mathsf{C}[\hat{P}_{ZW}\,\hat{P}_{XY}]    & \mathsf{C}[\hat{P}_{ZW}\,\hat{P}_{ZW}]  & \ldots   \\
    \vdots   & \vdots  & \ddots   \;
    \end{array}
    \right)\,.
\end{equation}

\subsection{Multipoles cross-correlation covariance}

\begin{figure}
    \centering
    \includegraphics[width=0.95\linewidth]{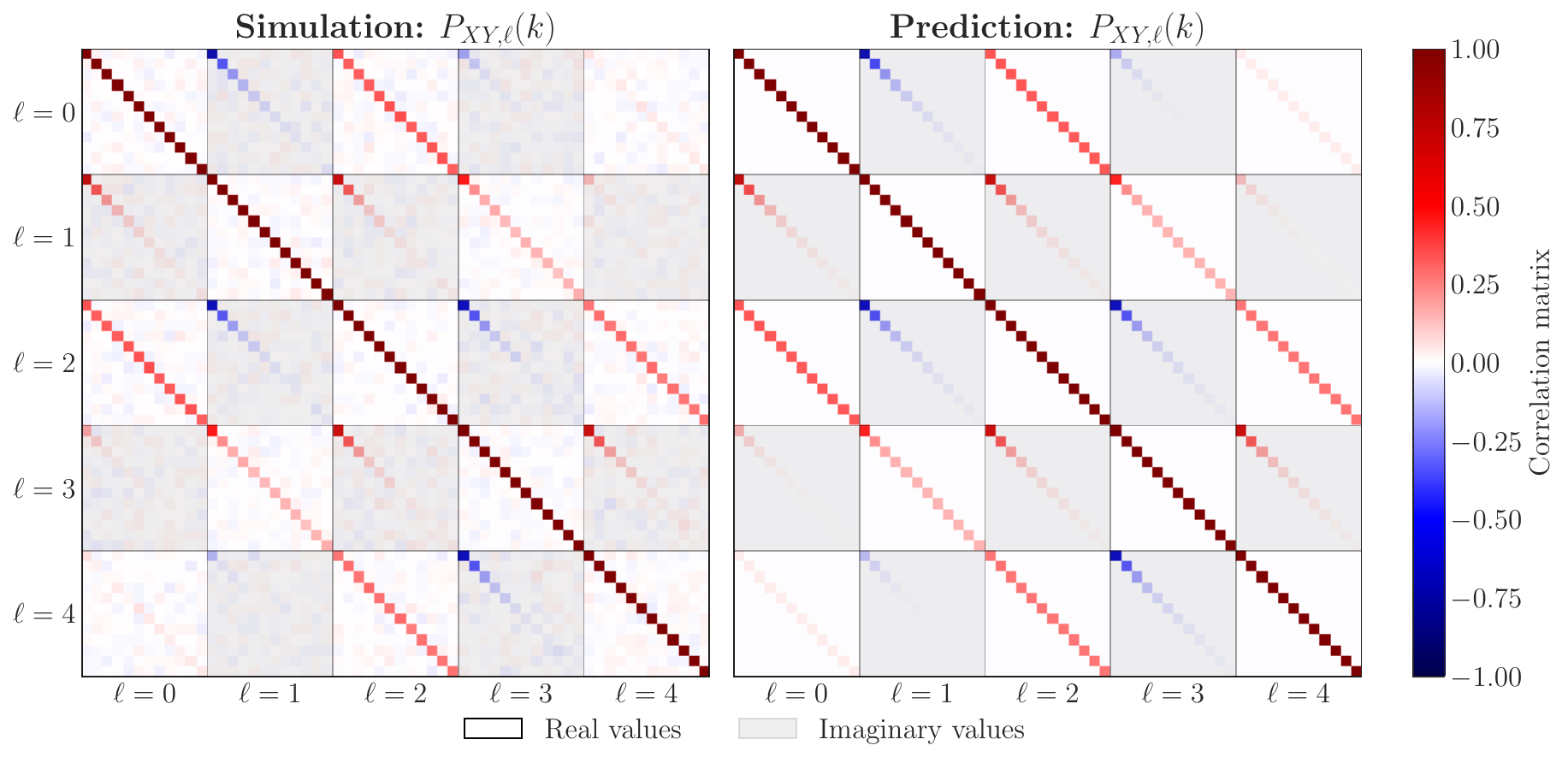}
    \caption{Correlation matrix associated with a measurement of the Legendre multipoles of the cross-power spectrum, $\hat{P}_{XY,\ell}(k)$. For comparison, the left panel shows the correlation matrix of our Gaussian realisations, while the right panel reports the theoretical prediction. The chessboard subdivision highlights the correlations between different multipoles, and is a result of having defined the data vector by concatenation of 5 (i.e.\ $\ell\in[0,5]$) sets of 12 (i.e.\ the number of bins in $k$) points. The correlation matrix is Hermitian, as can be appreciated from the white and grey tiles, respectively referring to real-only and imaginary-only values.}
    \label{fig:crr_cross_multipoles}
\end{figure}

In the two panels of \cref{fig:crr_cross_multipoles} we compare the cross-spectrum correlation matrix from the numerical realisations (left) and out theoretical prediction (right).
Both the measured cross-correlation matrix from the mock realisations and the corresponding analytical prediction of \cref{eq:covariance_Pl_full_multi} are arranged in a $\ell\times\ell'$ multipole-block layout with each block spanning the $k$-bin covariance between two multipole orders. 
By construction---namely, due to the definition of multipoles of the spectrum in Fourier space---even$\times$even and odd$\times$odd blocks are fully real, while even$\times$odd multipoles combinations are fully imaginary. This highlights the Hermitian structure of the covariance, which adds to the well-known symmetric structure a meaningful anti-symmetric pattern well-reproduced here.

\begin{figure}
    \centering
    \includegraphics[width=0.9\linewidth]{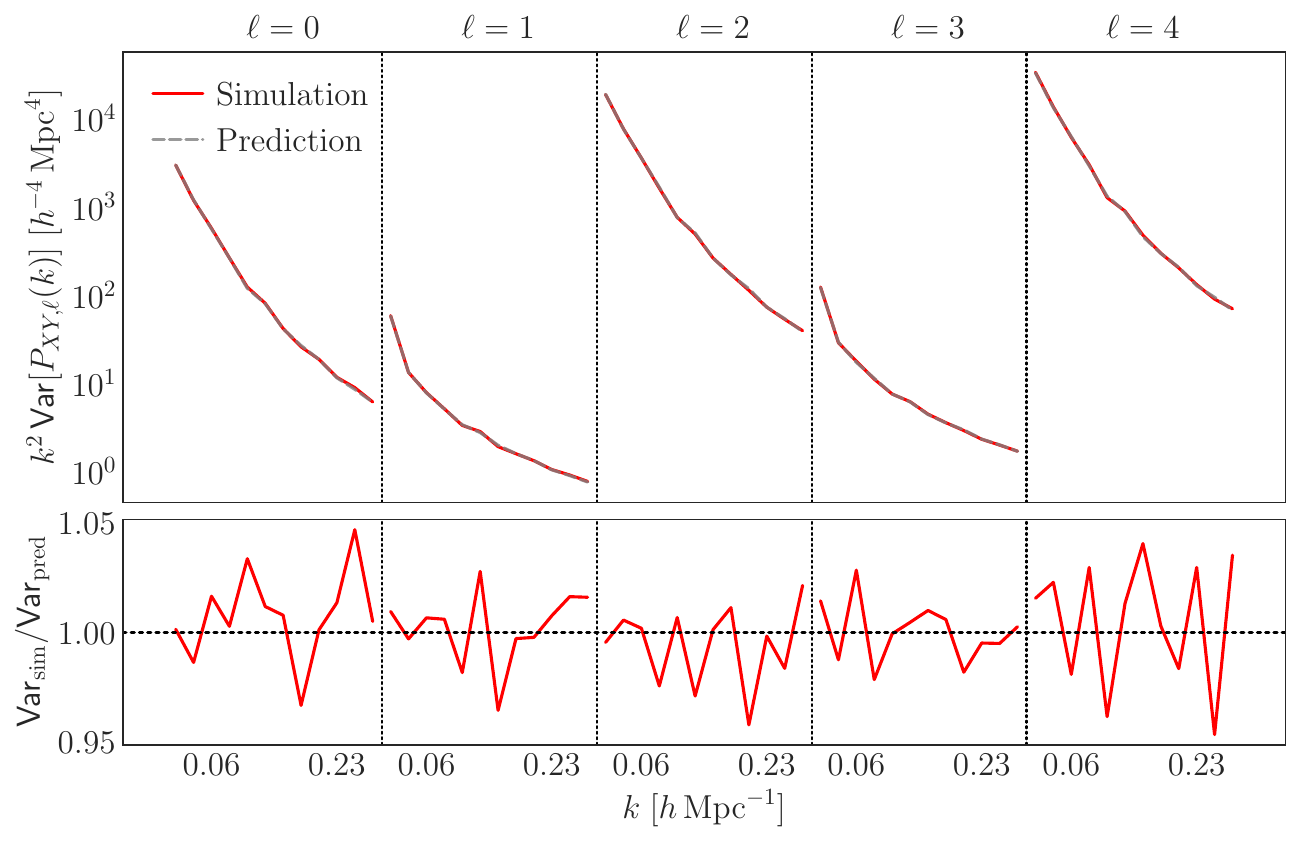}
    \caption{Diagonal elements of the covariance matrix of $\hat{P}_{XY,\ell}(k)$---that is, the variance of the measured cross-correlation power spectrum multipoles---for our simulations and analytical prediction. While the top panel depicts the variance itself, the lower one shows the ratio between the diagonal of the covariance matrix measured from the simulations and that of the theoretical prediction, hence stressing their mutual agreement.}
    \label{fig:var_cross_multipoles}
\end{figure}
The main diagonal of the matrices shown in \cref{fig:crr_cross_multipoles} is not informative, as it goes identically to unity for each correlation matrix by definition. For this reason, we compare in \cref{fig:var_cross_multipoles} the diagonal elements of the cross-multipole covariance matrix from simulations and the analytical prediction of \cref{eq:covariance_Pl_full_multi}.
The mutual agreement clearly does not show any evident feature signalling possible systematic errors.

\subsection{2D cross-power spectrum covariance} \label{sec:validation_pkmu}

\begin{figure}
    \centering
    \includegraphics[width=0.95\linewidth]{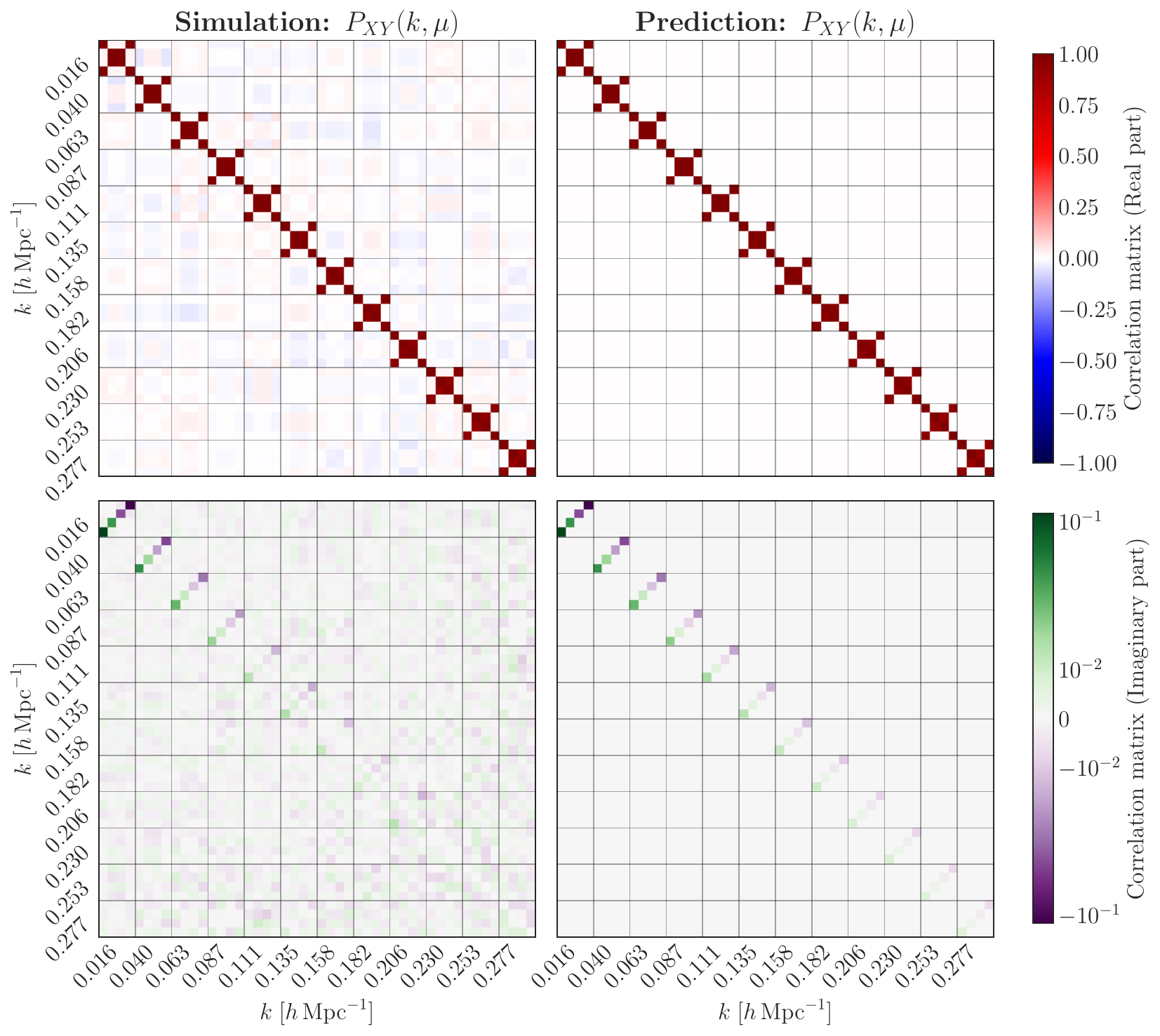}
    \caption{Similar to \cref{fig:crr_cross_multipoles}, but in the case of the 2D power spectrum. Real (top row) and imaginary (bottom row) parts of the correlation matrix associated with a measurement of $\hat{P}_{XY}(k,\mu)$ are separately reported, from simulations (left column) and analytical prediction (right column). The data vector is now ordered so that the index running over the 4 $\mu$-bins is faster than that on the 12 bins in $k$; as a result, every cell of the grid refers to a $(k,k')$ combination. Column-comparison tells about the agreement between theory and simulations; row-comparison informs about the Hermitian nature of the covariance matrix.}
    \label{fig:crr_cross_P2D}
\end{figure}
Concerning the 2D power spectrum, \cref{fig:crr_cross_P2D} presents a direct simulation-theory comparison of the $\hat{P}_{XY}(k,\mu)$ correlation matrix, organised in a four-panel layout that separates real and imaginary components. 
The top row shows the real-part correlation matrices from our Gaussian realisations and the analytical prediction of \cref{eq:cov_Pkmu_multi}, while the bottom row shows the corresponding imaginary parts, using a symmetric colour normalisation for the real component and a symmetrical logarithmic scaling for the weaker imaginary signal. The overlaid block grid marks transitions between $k$-bins, with each block containing the full structure in $\mu$, itself defined as 4 equispaced bins in $[-1,\,1]$. 
In analogy with \cref{fig:crr_cross_multipoles}, the imaginary part makes explicit the Hermitian nature of the covariance matrix.
A closer look to the anti-diagonal part of each $(k,k)$ cell reveals a satisfactory level of agreement between prediction and simulations of the off-diagonal terms, with discrepancies $<5\%$ at all scales in the real part, and at those where the signal-to-noise ratio is high in the imaginary part.

Moreover, an instructive exercise is looking at the real and imaginary parts of the cross-power spectrum, separately. In fact, estimates of the covariance matrix for a complex power spectrum can be derived directly from the theory of complex stochastic variables, where the full second-moment structure requires both the covariance and pseudo-covariance of the underlying complex random process. 
The theoretical variance of the real and imaginary parts of $\hat{P}_{XY}(k,\mu)$ then reads
\begin{align}
    \mathsf{Var}\left\{\Re[\hat{P}_{XY}(k,\mu)]\right\}=&\,\frac{1}{2}\, \Re \left\{\mathsf{C}[\hat{P}_{XY}(k,\mu),\,\hat{P}_{XY}(k,\mu)] + \mathsf{C}[\hat{P}_{XY}(k,\mu),\,\hat{P}_{YX}(k,\mu)]\right\}\;, \label{eq:var_PkmuXY_real} \\
     \mathsf{Var}\left\{\Im[\hat{P}_{XY}(k,\mu)]\right\}=&\,\frac{1}{2}\, \Re \left\{\mathsf{C}[\hat{P}_{XY}(k,\mu),\,\hat{P}_{XY}(k,\mu)] - \mathsf{C}[\hat{P}_{XY}(k,\mu),\,\hat{P}_{YX}(k,\mu)]\right\}\,, \label{eq:var_PkmuXY_imaginary}
\end{align}
where $\mathsf{C}[\hat{P}_{XY}(k,\mu),\,\hat{P}_{XY}(k,\mu)]$ represents the full variance and $\mathsf{C}[\hat{P}_{XY}(k,\mu),\,\hat{P}_{YX}(k,\mu)]=\mathsf{C}[\hat{P}_{XY}(k,\mu),\,\hat{P}_{XY}^\ast(k,\mu)]$ the so-called pseudo-variance; crucially, both quantities can be calculated with appropriate substitutions in \cref{eq:cov_Pkmu_multi}. Albeit \cref{eq:var_PkmuXY_real,eq:var_PkmuXY_imaginary} can be seen as application of the theory of complex random variables, we choose to leave in \cref{app:cov_imaginary_part} an in-depth derivation of the variance of the imaginary part of the cross-spectrum more in line with the approach of \cite{McDonald_2009}, who first addressed this calculation in the context of galaxy clustering analyses.

\begin{figure}
    \centering
    \includegraphics[width=0.95\linewidth]{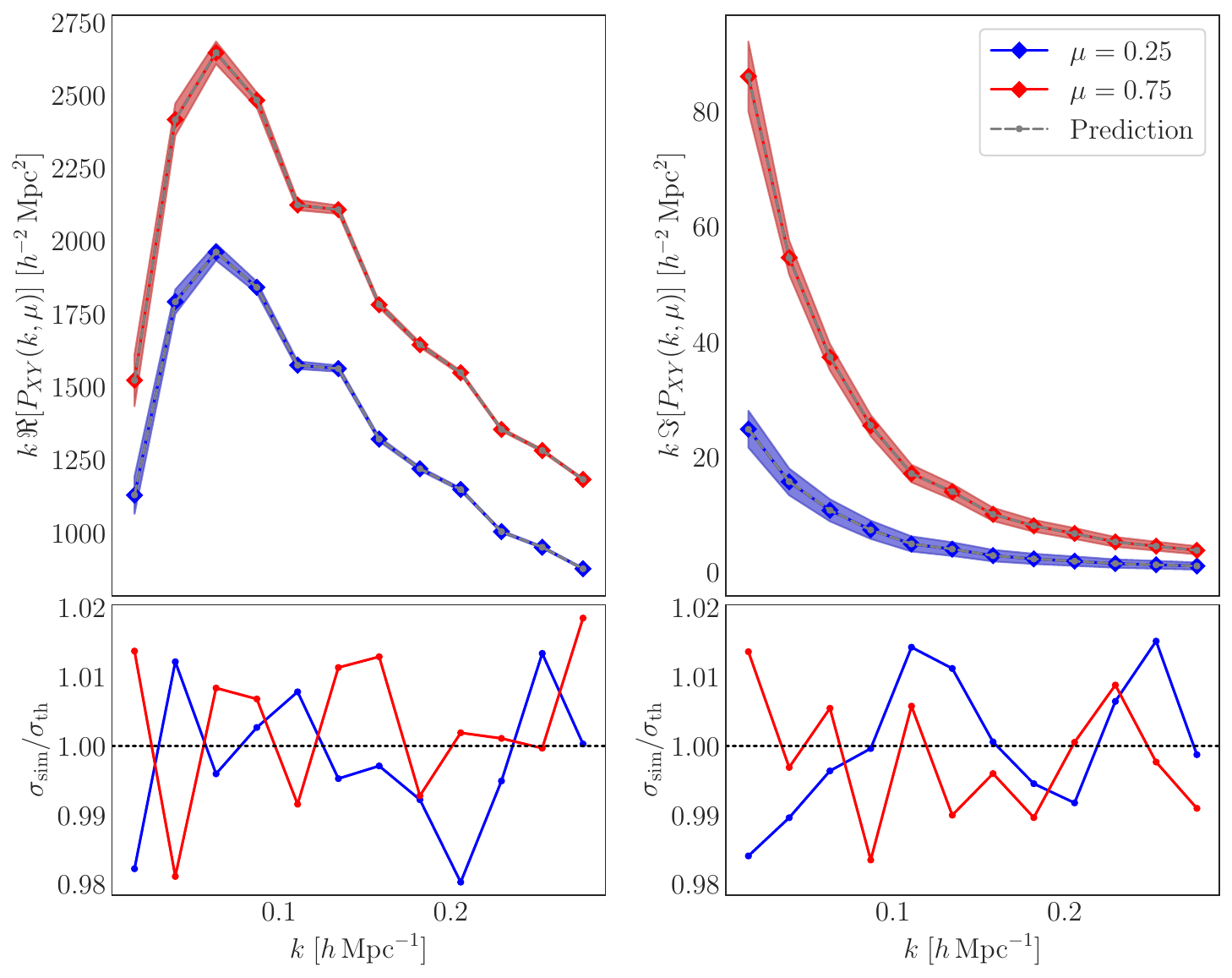}
    \caption{Comparison between measured cross-2D power spectrum and analytical predictions. Left and right panels display the real and imaginary parts of the spectrum, respectively; the top panels give the (simulated and predicted) $\hat{P}_{XY}(k,\mu)$, whilst the bottom panels present the striking agreement between the predicted standard deviation and that extracted from our realisations, showing their ratio. The same binning as in \cref{fig:crr_cross_P2D} is adopted, but only two bins in $\mu$ ($\mu >0$, marked in different colours) are shown since the remaining ones can be inferred by symmetry(asymmetry) arguments in the real(imaginary) signal.}
    \label{fig:crr_cross_realimag_P2D}
\end{figure}
\Cref{fig:crr_cross_realimag_P2D} compares the simulated and predicted cross-correlation 2D power spectrum by separating its real and imaginary components and displaying both their amplitudes and associated uncertainties. 
The upper panels show $k\,\Re[P_{XY}(k,\mu)]$ and $k\,\Im[P_{XY}(k,\mu)]$. Here, coloured curves and shaded bands denote the mean and $1\sigma$ scatter measured from the Gaussian simulations, while the dashed gray curves and lighter gray bands depict the corresponding theoretical predictions and uncertainties.
The lower panels quantify the good agreement at the level of the standard deviation through the ratio $\sigma_{\rm sim}/\sigma_{\rm th}$, indicating clear consistency between simulated and predicted variances.

Our 2D power spectrum also offers the possibility to infer the covariance of clustering wedges, where the $\mu$ and $-\mu$ bins are always coupled together. Interpreting them as pairs of bins in ours \cref{eq:estPkmu,eq:cov_Pkmu_multi}, two aspects become clear. First, clustering wedges will have a real data vector by construction, since imaginary---thus of odd-parity---contributions to $P_{XY}(\qv)$ average to zero in a $|\mu|$-bin, with $-\mu-\Delta\mu/2\le \mu_{\qv} < -\mu+\Delta\mu/2$ {\em or}  $\mu-\Delta\mu/2\le \mu_{\qv} < \mu+\Delta\mu/2$. Second, the associate covariance matrix will be diagonal---with no correlations among different bins---hence a real, symmetric quantity. Each block of the kind of \cref{eq:cov_Pkmu_block_cross} (appreciable in \cref{fig:crr_cross_P2D} too) will collapse into a diagonal element of $\mathsf{C}[\hat{P}_{XY}(k,|\mu|),\,\hat{P}_{XY}(k',|\mu'|)]$, which in turn will produce a well-defined matrix also for auto-correlation power spectra.

\section{Conclusion}
\label{sec:conclusion}

In this work, we have presented a generalisation of existing Gaussian covariance expressions to a two-point multi-tracer scenario, encompassing both real and complex power spectra. 
Imaginary, odd-parity contributions emerge in clustering summary statistics as a prediction of general relativity on cosmological scales \cite{McDonald_2009,Jeong:2019igb,2024_Paul_PhRvL}. With their first detection foreseen thanks to upcoming data, their inclusion in cosmological analyses appears both timely and compelling.
For this reason, our general results apply both to real---that is, standard, even-parity only---and complex power spectra. 
In addition, joint auto- and cross-spectrum fits are expected to deliver unprecedented constraints on large-scale physics by limiting cosmic variance \cite{2009_Seljak_PhRvL,2009_McDonald&Seljak_JCAP,2013_Abramo&Leonard_MNRAS}. Therefore, the formulae we have reported consider the broadest possible multi-tracer scenario for two-point statistics: they allow contributions to the covariance matrix from up to four different tracers, and no cross-covariance between fully independent cross-power spectra can have more.

In \cref{sec:covariance} we have outlined our main results. For a generic, weighted estimator, the multi-tracer covariance is given by \cref{eq:covPs_general}, and then rephrased in the useful cases of the auto- and cross-power spectrum variances. 
Building on previous works \cite{Smith_2009,Grieb_2016}, we have then retrieved the covariance matrix of the Legendre multipoles of the power spectrum, $\hat{P}_{XY,\ell}(k)$, and the 2D power spectrum, $\hat{P}_{XY}(k,\mu)$, commenting on how the well-known limits are coherently recovered via further simplifications (e.g.\ real data vector, monopole, clustering wedges, etc.). 
\Cref{sec:validation} guides through an in-depth understanding of the structure and properties of the Gaussian covariance matrix, while comparing the accuracy of our predictions against Gaussian simulations, with specific interest in the Hermitian behaviour of its imaginary part. 

The assumptions of Gaussianity and trivial survey geometry have been held throughout our derivations. We defer to future developments of this work aspects that can be incorporated in the framework developed in Ref.\ \cite{Wadekar_2020}: the consideration of non-Gaussian contributions to the covariance matrix and convolution with the survey window, as well as the issue of overcoming the plane-parallel approximation. 

Our work's originality lies in the extension of the calculation of the power spectrum Gaussian covariance matrix to a general multi-tracer scenario, while accounting for odd-parity components in the data vector. This has a relevant impact, as explorative studies that make use of cross-correlations and/or focus on odd-parity effects in the linear regime need an appropriate covariance matrix, consistent among tracers and fully Hermitian. 

\acknowledgments
FM and SC are grateful to L.R.\ Abramo for useful discussions in the early stages of this work. They also thank C.\ Clarkson and R.\ Maartens for reading the final version of the draft.
FM and SC acknowledge support from the Italian Ministry of University and Research (\textsc{mur}), PRIN 2022 `EXSKALIBUR – Euclid-Cross-SKA: Likelihood Inference Building for Universe's Research', Grant No.\ 20222BBYB9, CUP D53D2300252 0006, and from the European Union -- Next Generation EU. MYE acknowledges support by the PRIN 2022 PNRR project ``Space-based cosmology with Euclid: the role of High-Performance Computing" (code no. P202259YAF), funded by European Union – Next Generation EU”. ES acknowledges support from the Theory Grant 2023 “NeuMass” of the Italian National Institute for Astrophysics (INAF).


\clearpage
\appendix

\section{Derivation of power spectrum multipoles covariance} \label{app:derivation_covPell}
Here we provide the full derivation of \cref{eq:covariance_Pl_full_multi}, starting from \cref{eeq:cov_Pell_start}.
In summary, we translate \cref{eeq:cov_Pell_start} into the infinite volume limit with the substitution $\sum_{\qv \in k} \longrightarrow k_{\rm f}^{-3}\int_k$; then, we expand in multipoles the power spectra that appear in the sum over the modes in the bin $k$, using \cref{eq:multipole_exp}; and we deal with the integration of the Legendre polynomials over $\mu_{\qv}$, i.e.\ the angular integration.

We begin by working out a useful relation. Thanks to the identity
\begin{equation}
    \mathcal{L}_{\ell_1}(\mu) \mathcal{L}_{\ell_2}(\mu) = \sum_{\ell_3} (2\,\ell_3 + 1) 
    \begin{pmatrix} 
        \ell_1 & \ell_2 & \ell_3 \\ 
        0 & 0 & 0 
    \end{pmatrix}^2
    \mathcal{L}_{\ell_3}(\mu)\,,
\end{equation}
and the orthogonally of Legendre polynomials, 
\begin{equation}
    \int_{-1}^1 \de \mu \, \mathcal{L}_{\ell_1}(\mu) \, \mathcal{L}_{\ell_2}(\mu)=\frac{2\,\delta^{\rm K}_{\ell_1,\ell_2}}{2\,\ell_1+1}\,,
\end{equation}
we calculate
\begin{multline}
     \int_{-1}^1 \de \mu \, \mathcal{L}_{\ell_1}(\mu) \, \mathcal{L}_{\ell_2}(\mu) \, \mathcal{L}_{\ell_3}(\mu) \, \mathcal{L}_{\ell_4}(\mu) = \,  \int_{-1}^1 \de \mu \, \left[\sum_{L_1} (2\,L_1 + 1) 
    \begin{pmatrix} 
        \ell_1 & \ell_2 & L_1 \\ 
        0 & 0 & 0 
    \end{pmatrix}^2
    \mathcal{L}_{L_1}(\mu) \right] \\
    \times \left[\sum_{L_2} (2\,L_2 + 1) 
    \begin{pmatrix} 
        \ell_3 & \ell_4 & L_2 \\ 
        0 & 0 & 0 
    \end{pmatrix}^2
    \mathcal{L}_{L_2}(\mu) \right] \\
     =\sum_{L_1,L_2} \,  (2\,L_1 + 1) \, (2\,L_2 + 1) 
    \begin{pmatrix} 
        \ell_1 & \ell_2 & L_1 \\ 
        0 & 0 & 0 
    \end{pmatrix}^2
    \begin{pmatrix} 
        \ell_3 & \ell_4 & L_2 \\ 
        0 & 0 & 0 
    \end{pmatrix}^2
    \\
    \times \int_{-1}^1 \de \mu \, \mathcal{L}_{L_1}(\mu) \,\mathcal{L}_{L_2}(\mu) \,\sum_{L_1} \, 2 \,  (2\,L_1 + 1) 
    \begin{pmatrix} 
        \ell_1 & \ell_2 & L_1 \\ 
        0 & 0 & 0 
    \end{pmatrix}^2
    \begin{pmatrix} 
        \ell_3 & \ell_4 & L_1 \\ 
        0 & 0 & 0 
    \end{pmatrix}^2\,. \label{eq:4Legendre_int}
\end{multline}
We point out that \cref{eq:4Legendre_int} is not in agreement with eq.\ (A4) in Ref.\ \cite{Grieb_2016}, which lacks a $(2\,L_1 + 1)$ factor.  

At this point, we write \cref{eeq:cov_Pell_start} in the continuum limit,
\begin{align}
    \mathsf{C}[\hat{P}_{XY,\,\ell}(k),\,\hat{P}_{ZW,\,\ell'}(k')  ] &= 
    \frac{(2\,\ell+1)\,(2\,\ell'+1)\,\delta^{\rm K}_{k,k'}}{N_k^2} 
    \sum_{\qv \in k} \,\Le_\ell(\mu_{\qv}) \, \Le_{\ell'}(\mu_{\qv})
    \nonumber \\&\quad
    \times\;\left[P_{XZ}(\qv)\,P_{WY}(\qv)
    + (-1)^{\ell'}\, P_{XW}(\qv)\,P_{ZY}(\qv)\right] \\
    & =  \frac{(2\,\ell+1)\,(2\,\ell'+1)\,\delta^{\rm K}_{k,k'}}{N_k^2\, k_{\rm f}^3}
    \int_k \de^3 \qv\,
    \Le_\ell(\mu_{\qv}) \, \Le_{\ell'}(\mu_{\qv}) \nonumber\\
    &\quad \times\;
     \left[P_{XZ}(\qv) \, P_{WY}(\qv) \,+ 
     (-1)^{\ell'}\,P_{XW}(\qv) \, P_{ZY}(\qv)\right]\,, \label{eq:multi_cov_der1}
\end{align}
and insert \cref{eq:multipole_exp} into \cref{eq:multi_cov_der1},
\begin{align}
    \mathsf{C}[\hat{P}_{XY,\,\ell}(k),\,\hat{P}_{ZW,\,\ell'}(k')  ] 
    & =  \frac{2\,\pi\,(2\,\ell+1)\,(2\,\ell'+1)\,\delta^{\rm K}_{k,k'}}{N_k^2\, k_{\rm f}^3}
    \int_{k-\Delta k/2}^{k+\Delta k/2} \de q\, q^2 \int_{-1}^1 \de \mu_{\qv} \,
    \Le_\ell(\mu_{\qv}) \, \Le_{\ell'}(\mu_{\qv}) \nonumber\\
    &\quad \times\;
     \Bigg[ \Bigg(\sum _{L_1}P_{XZ,L_1}(q)\Le_{L_1}(\mu_{\qv}) \Bigg)  \Bigg( \sum_{L_2} P_{WY,L_2}(q) \Le_{L_2}(\mu_{\qv})\Bigg)  \nonumber \\
    &\quad \quad + (-1)^{\ell'}\,\Bigg(\sum _{L_3}P_{XW,L_3}(q)\Le_{L_3}(\mu_{\qv}) \Bigg) \Bigg( \sum_{L_4} P_{ZY,L_4}(q) \Le_{L_4}(\mu_{\qv})\Bigg) \Bigg] \\   
    & = \frac{2\,\pi\,(2\,\ell+1)\,(2\,\ell'+1)\,\delta^{\rm K}_{k,k'}}{N_k^2\, k_{\rm f}^3}
    \int_{k-\Delta k/2}^{k+\Delta k/2} \de q\, q^2 \int_{-1}^1 \de \mu_{\qv} \,
    \Le_\ell(\mu_{\qv}) \, \Le_{\ell'}(\mu_{\qv}) \nonumber\\
    &\quad \times\;
     \Bigg\{ \sum _{L_1,L_2}P_{XZ,L_1}(q) \, P_{WY,L_2}(q) \, \Le_{L_1}(\mu_{\qv}) \,\Le_{L_2}(\mu_{\qv})  \nonumber \\
    &\quad \quad + (-1)^{\ell'}\,\sum _{L_3,L_4}P_{XW,L_3}(q)\, P_{ZY,L_4}(q) \,\Le_{L_3}(\mu_{\qv}) \, \Le_{L_4}(\mu_{\qv}) \Bigg\} \\
    & =  \frac{2\,\pi\,(2\,\ell+1)\,(2\,\ell'+1)\,\delta^{\rm K}_{k,k'}}{N_k^2\, k_{\rm f}^3} \sum _{L_1,L_2}\, 
    \int_{k-\Delta k/2}^{k+\Delta k/2} \de q\, q^2  \nonumber\\
    &\quad \times\;
    \Big\{P_{XZ,L_1}(q) \, P_{WY,L_2}(q) + (-1)^{\ell'}\, P_{XW,L_1}(q)\, P_{ZY,L_2}(q) \Big\} \nonumber \\
    & \quad \quad \times \int_{-1}^1 \de \mu_{\qv} \,
    \Le_\ell(\mu_{\qv}) \, \Le_{\ell'}(\mu_{\qv}) \,  \Le_{L_1}(\mu_{\qv}) \,\Le_{L_2}(\mu_{\qv}) \,,
\end{align}
where we have exploited the fact that the expansions over $L_3$ and $L_4$ run over the same indices as those on $L_1$ and $L_2$. Finally, we utilise \cref{eq:4Legendre_int} and find
\begin{align}
    \mathrm{Cov}({\hat P_{XY,\ell}}({k}),&\, {\hat P_{ZW,\ell'}}({ k'})) = \nonumber \\
    & 
    \frac{2\,\pi\,(2\,\ell+1)\,(2\,\ell'+1)\,\delta^{\rm K}_{k,k'}}{N^2_k \, k_{\rm f}^3} \,  
    \sum _{L_1,L_2,L_3} \, (2\,L_3+1) \,
    \begin{pmatrix} 
        \ell & \ell' & L_3 \\ 
        0 & 0 & 0 
    \end{pmatrix}^2
    \begin{pmatrix} 
        L_1 & L_2 & L_3 \\ 
        0 & 0 & 0 
    \end{pmatrix}^2
    \nonumber\\
    &\qquad 
    \int_{k-\Delta k/2}^{k+\Delta k/2}
    \de q\, q^2\,\Big\{P_{XZ,L_1}(q) \, P_{WY,L_2}(q) + (-1)^{\ell'}\, P_{XW,L_1}(q)\, P_{ZY,L_2}(q) \Big\} \,,
\end{align}
which is equivalent to \cref{eq:covariance_Pl_full_multi}.

\section{Covariance of the imaginary part of the spectrum} \label{app:cov_imaginary_part}
A fairly insightful exercise is taking a closer look at the imaginary part of the cross-spectrum, which is expected to display a signal due to relativistic projection effects. 
Given a theoretical cross-correlation power spectrum, the imaginary part can be extracted out of a convenient combination of the power spectra themselves,
\begin{equation} \label{eq:imaginary_cross_def}
    \Im[P_{XY}(\qv)] = -\frac\ii2\,P_{[XY]}(\qv)\;,
\end{equation}
where square brackets around indices denote anti-symmetrisation.
In line with Ref.\ \cite{McDonald_2009}, the covariance matrix of any of its estimators will thus imply the computation of
\begin{multline}
    \langle \Im [P_{XY}(\qv)]\, \Im [P_{XY}({\qv'})] \rangle - \langle \Im [P_{XY}(\qv)]\rangle\,\langle \Im [P_{XY}(\qv')] \rangle = \\ \underbrace{-\frac{1}{4}\,\langle[\delta_X(\qv)\,\delta_Y(-\qv)-\delta_X(-\qv)\,\delta_Y(\qv)]\,[\delta_X({\qv'})\,\delta_Y({-\qv'})-\delta_X({-\qv'})\,\delta_Y({\qv'})] \rangle}_{L} \\
    + \underbrace{\frac{1}{4}\,\langle\delta_X(\qv)\,\delta_Y(-\qv)-\delta_X(-\qv)\,\delta_Y(\qv) \rangle \,\langle \delta_X({\qv'})\,\delta_Y({-\qv'})-\delta_X({-\qv'})\,\delta_Y({\qv'}) \rangle}_{K}\,, \label{eq:cov_imaginarypart}
\end{multline}
with no complex conjugation of the second term, since $\Im[P_{XY}(\qv)]$ is defined as a real quantity. Above, $L$ gives
\begin{align}
    L
    &= -\frac{1}{4}\,[\langle\delta_X(\qv)\,\delta_Y(-\qv)\,\delta_X({\qv'})\,\delta_Y({-\qv'}) \rangle +\langle\delta_X(-\qv)\,\delta_Y(\qv)\,\delta_X({-\qv'})\,\delta_Y({\qv'}) \rangle \nonumber \\
    &\qquad -\langle\delta_X(\qv)\,\delta_Y(-\qv)\,\delta_X({-\qv'})\,\delta_Y({\qv'}) \rangle -\langle\delta_X(-\qv)\,\delta_Y(\qv)\,\delta_X({\qv'})\,\delta_Y({-\qv'}) \rangle]  \\
    &= -\frac{1}{2}\,\left\{ \Re[\langle\delta_X(\qv)\,\delta_Y(-\qv)\,\delta_X({\qv'})\,\delta_Y({-\qv'}) \rangle]  -\Re[\langle\delta_X(\qv)\,\delta_Y(-\qv)\,\delta_X({-\qv'})\,\delta_Y({\qv'}) \rangle]\right\}  \\
    &= -K -\frac{1}{2}\,\big\{ \Re[\langle\delta_X(\qv)\,\delta_X({\qv'})\rangle\,\langle\delta_Y(-\qv)\,\delta_Y({-\qv'}) \rangle] + \Re[\langle\delta_X(\qv)\,\delta_Y({-\qv'})\rangle\,\langle\delta_Y(-\qv)\,\delta_X(\qv') \rangle] \nonumber \\
    &\qquad \qquad -\Re[\langle\delta_X(\qv)\,\delta_X({-\qv'})\rangle\,\langle\delta_Y(-\qv)\,\delta_Y({\qv'}) \rangle] - \Re[\langle\delta_X(\qv)\,\delta_Y({\qv'})\rangle\,\langle\delta_Y(-\qv)\,\delta_X({-\qv'}) \rangle]\big\}\,; \label{eq;imag_term1}
\end{align}
where we have used the realty condition to extract the real parts and Wick's theorem again to simplify the four-point functions. We also notice that the first terms of that expansion coincide with $K$ in \cref{eq:cov_imaginarypart}. 
Substituting \cref{eq;imag_term1} in \cref{eq:cov_imaginarypart}, we then have
\begin{multline}
    \langle \Im [P_{XY}(\qv)]\,\Im [P_{XY}({\qv'})] \rangle -\, \langle \Im [P_{XY}(\qv)]\rangle\,\langle \Im [P_{XY}(\qv')] \rangle = \\ 
    \frac{(2\,\pi)^6}{2\,k_{\rm f}^6} \, \left\{\Re [P_{XX}(\qv) \, P_{YY}(\qv)]-\Re [P_{XY}^2(\qv)]\right\}  \, \left(\, \delta^{\rm K}_{\qv,\qv'} \, \delta^{\rm K}_{\qv,\qv'}-\, \delta^{\rm K}_{\qv,-\qv'} \, \delta^{\rm K}_{\qv,-\qv'}\right).\label{eq:Cov_modebymode_imaginarypart}
\end{multline}

It is worth noticing that \cref{eq:Cov_modebymode_imaginarypart} is in agreement with Ref.\ \cite{McDonald_2009}. Specifically, Eqs\ (2.15) and (2.16) in Ref.\ \cite{McDonald_2009} are the mode-by-mode variance of the imaginary part of the cross-spectrum $P_{XY}$; that is, only the diagonal part in $\qv$ (the first term in the round brackets of \cref{eq:Cov_modebymode_imaginarypart}).
Interestingly, the variance factor in \cref{eq:Cov_modebymode_imaginarypart} is equivalent to the prediction of \cref{eq:var_PkmuXY_imaginary} (as anticipated in \cref{sec:validation_pkmu}) therefore this derivation can be seen as a sort of proof of its validity.

Assuming a linear regime and dubbing respectively $R_X$ and $I_X$ the real and the imaginary parts of an effective transfer function that multiples the matter density contrast $\delta_{\rm m}$---so that $\delta_X(\qv)\coloneqq[R_X(\qv)+\ii\,I_X(\qv)]\, \delta_{\rm m}(\qv)$ and $\Im[P_{XY}(\bm k)]=R_{[Y}(\bm k)\,I_{X]}(\bm k)\,P_{\rm m}(k)$, with $P_{\rm m}$ the matter power spectrum---we are able to simplify (for clarity, let us omit the $\qv$ dependence in this case)
\begin{align}
    \Re [P_{XX} \, P_{YY}]-\Re [P_{XY}^2]=&\,(R_X^2+I_X^2)\,(R_Y^2+I_Y^2)-\Re[R_X\,R_Y+I_X\,I_Y+\ii\,(R_Y\,I_X-R_X\,I_Y)]^2 \nonumber\\ &\qquad \times  P_{\rm m} \label{eq:imaginarypart_finalvariance1} \\
    =&\, 
        (R_X^2+I_X^2)\,(R_Y^2+I_Y^2)-(R_X\,R_Y+I_X\,I_Y)^2+(R_Y\,I_X-R_X\,I_Y)^2 \nonumber\\ &\qquad \times  P_{\rm m}\nonumber \\
    =&\,
        R_X^2\,I_Y^2+R_Y^2\,I_X^2-2\,R_X\,R_Y\,I_X\,I_Y+(R_Y\,I_X-R_X\,I_Y)^2\, P_{\rm m} \nonumber \\
    =&\,
        2\,(R_Y\,I_X-R_X\,I_Y)^2\, P_{\rm m} \nonumber \\
    =&\,
        2\,\left(\Im[P_{XY}]\right)^2\,, \label{eq:imaginarypart_finalvariance2} 
\end{align}
and obtain the ``without-noise" expression of eq. (2.15) in Ref.\ \cite{McDonald_2009} (see also Appendix B in Ref.\ \cite{Addis_2025}). 
On the other hand, including noise means applying the substitution (\ref{eq:noise}) in \cref{eq:imaginarypart_finalvariance1}. This gives us extra terms---i.e.\ the mixed spectrum-noise and the noise-noise factors---in the left-hand side, which remain unaffected by the subtraction of $\Re [P_{XY}^2]$. \Cref{eq:imaginarypart_finalvariance2} thus becomes
\begin{equation}
      \Re [P_{XX} \, P_{YY}]-\Re [P_{XY}^2]=2\,\left(\Im[P_{XY}]\right)^2+\left(\frac{P_{XX}}{\bar n_Y}+\frac{P_{YY}}{\bar n_X}+\frac{1}{\bar n_X \bar n_y}\right)\,,
\end{equation}
in consistency with eq.\ (2.16) of \cite{McDonald_2009}.

\section{Validation details} \label{app:validation_details}
Intending to make our validation method replicable, we report in \cref{tab:values} the tracer-dependent quantities (in \crefnp{eq:T_delta_g_Fourier_space}) we have assumed throughout \cref{sec:validation}. 
A flat $\Lambda$CDM \cite{2018_Planck_results} cosmology is used, and the spectra are evaluated at $z=0.1$.
As stated before, the accurate modelling of the properties of the LSS tracers, as well as them being galaxy samples, is not a concern for our numerical test of the correctness of the analytical derivation.

\begin{table} [h]
    \centering
    \begin{tabular}{c|cc}
       Tracer:         & $X$       & $Y$       \\
       \hline 
       $b$      & $1.5$     & $2.2$     \\
       $\alpha$ & $10.0$    & $-10.0$   \\
       $\bar n \;[10^{-2}\, h^3 \rm{Mpc}^{-3}]$ & $1.0$     & $0.1$     \\
    \end{tabular}
    \caption{Assumed values for redshift-dependent parameters of our Gaussian simulations. They are not meant to reproduce any specific galaxy sample.}
    \label{tab:values}
\end{table}


\bibliographystyle{jhep}
\bibliography{bibliography}


\end{document}